\documentclass[prl,reprint,twocolumn,showpacs,preprintnumbers,amsmath,amssymb,floatfix]{revtex4}
\usepackage[german,english]{babel}
\usepackage{bm,amsmath,amsfonts}
\usepackage{amssymb}
\usepackage{graphicx}
\usepackage{xcolor}

\def\beq{\begin{equation}}
\def\eeq{\end{equation}}
\def\beqn{\begin{eqnarray}}
\def\eeqn{\end{eqnarray}}

\begin{document}

\title{Interferometry with correlated matter-waves}

\author{Oksana I. Streltsova$^1$ and Alexej I. Streltsov$^2$}

\affiliation{$^1$
Laboratory of Information Technologies,
Joint Institute for Nuclear Research,
Joliot-Curie 6, Dubna, Russia}

\affiliation{$^2$ Theoretische Chemie, Physikalisch-Chemisches Institut, Universit\"at Heidelberg, INF 229, Germany}

\begin{abstract}

Matter-wave interferometry of ultra-cold atoms with attractive interactions is studied at the full many-body level.
First, we study how a coherent light-pulse applied to an initially-condensed solitonic system splits it into two matter-waves.
The split system looses its coherence and develops correlations with time, and inevitably becomes fragmented due to inter-particle attractions.
Next, we show that by re-colliding the sub-clouds constituting the split density together, along with a simultaneous application of the same laser-pulse,
one creates three matter-waves propagating with different momenta. 
We demonstrate that the number of atoms in the sub-cloud with zero-momentum is directly proportional to 
the degree of fragmentation in the system.
This interferometric-based protocol to discriminate, probe, and measure the fragmentation is general and can be applied 
to ultra-cold systems with attractive, repulsive, short- and long-range interactions.

\end{abstract}

\pacs{03.75.Kk, 05.30.Jp, 03.65.-w, 03.75.Hh}

\maketitle

Interferometry with ultra-cold atomic beams increases its popularity
due to offered advantages on sensitivity, control, and accessibility of the quantum limit in resolution \cite{Interferometry_Rev}.
Recently, low-dimensional Bose-condensed systems with attractive inter-particle 
interactions become of special interest  \cite{Interferometry_Soliton_exp1,Interferometry_Soliton_exp2}
because the widths of wave-packets in these systems are stabilized owning to the underlying solitonic nature \cite{exp1,exp2,exp3,exp4,exp5},
which is expected to improve the resolution and sensitivity of matter-wave-based interferometers further on. 
The most undesired phenomenon in these interferometers is the degradation of coherence,
manifesting itself in a loss of visibility of interference fringes.
In matter-wave experiments it originates to the inter-particle interactions
and can be caused by quantum depletion, incoherent excitations of atoms due to temperatures, particle losses, etc.,
resulting in occupations of very many different quantum modes,
or by a development of fragmentation -- macroscopic occupation of a few modes.  
The fragmentation phenomena being extensively studied theoretically \cite{frag0, frag1,frag2,frag3,frag4,frag5} 
still have open questions.
The most basic and intriguing among them is:
How to measure fragmentation directly and which experimentally-accessible 
quantity to use for it?
Experimental verification of the fragmentation has been done so far only in ultra-cold atomic systems 
with repulsive interactions \cite{Kasevich0,JoergNatPhys06}.
In attractive systems the theoretical understanding of quantum depletion and fragmentation 
at the many-body level is limited \cite{MB_att1,fragmenton,MB_att3, swift,MB_att5,MB_att6,MB_att7},
especially in the interferometry-related context. 
Here, the majority of investigations and analysis \cite{Interferometry_Soliton_exp2,exp1,exp2,exp3,exp4,exp5} 
have been done at the mean-field level within the popular Gross-Pivaevskii (GP) theory \cite{Pitaevskii_review,Leggett_review}.
The GP equation, having intrinsic non-linear structure, equips us with bright solitonic solutions
whose densities have a single-  and multi-hump structures,
but is incapable for describing incoherence and fragmentation therein.

In this Letter we investigate the interferometry of attractive ultra-cold systems at the many-body level,
which allows us to trace the degradation of coherence and development of fragmentation between the split matter-waves.
As a result, we  show how a simple protocol, implementable within currently-available experimental facilities,
can be used for the direct measurement of fragmentation in general quantum systems with multi-hump structures in their densities.

A typical interferometric experimental sequence involves three stages: first -- to split the original beam into two parts,
second -- to deflect the parts propagating along different paths, and final -- to recombine them back together and 
to measure the resulting interference pattern.
In the context of ultra-cold atoms, the stage of  splitting the original 
initially-coherent cloud (beam) into the sub-clouds seems to be the most challenging one.
It can be done either by ramping-up a potential barrier,
or by applying additional laser pulses of special frequencies and configurations, transferring momentum to the atoms,
and separating thereby the sub-clouds (i.e., interferometers arms) in momentum space.
The process of splitting of Bose-Einstein condensate (BEC) by ramping-up a potential barrier is known \cite{PRAZoller2001,ramp_up} to be very non-trivial 
due to the accompanying loss of coherence and development of fragmentation in the final states.
In contrast to it, the momentum-transferred separations
induced by laser-splitters based on, e.g., Raman \cite{Kasevich2}, Bragg, or Kapitza-Dirac techniques \cite{Interferometry_Rev},
are believed to be more {\it coherent}.

We challenge the correctness of this assertion in low-dimensional BECs with {\it attractive} inter-particle interactions
by straightforward simulation of all the necessary interferometric steps via the solution of the time-dependent many-boson Schr\"odinger equation (TDSE)
$\hat H \Psi\!=\!i\hbar \frac{\partial\Psi}{\partial t}$. We clarify the role of many-body physics
by contrasting the TDSE many-body results and the Gross-Pitaevskii  (GP) mean-field predictions.
To solve the TDSE we use the multiconfigurational time-dependent Hartree for bosons (MCTDHB) method \cite{ramp_up,MCTDHB_paper,MCTDHB_HIM},
see, e.g., the books \cite{book1,book2}.

\begin{figure}[t]
\vspace*{0cm}
\includegraphics[width=8cm,angle=0] {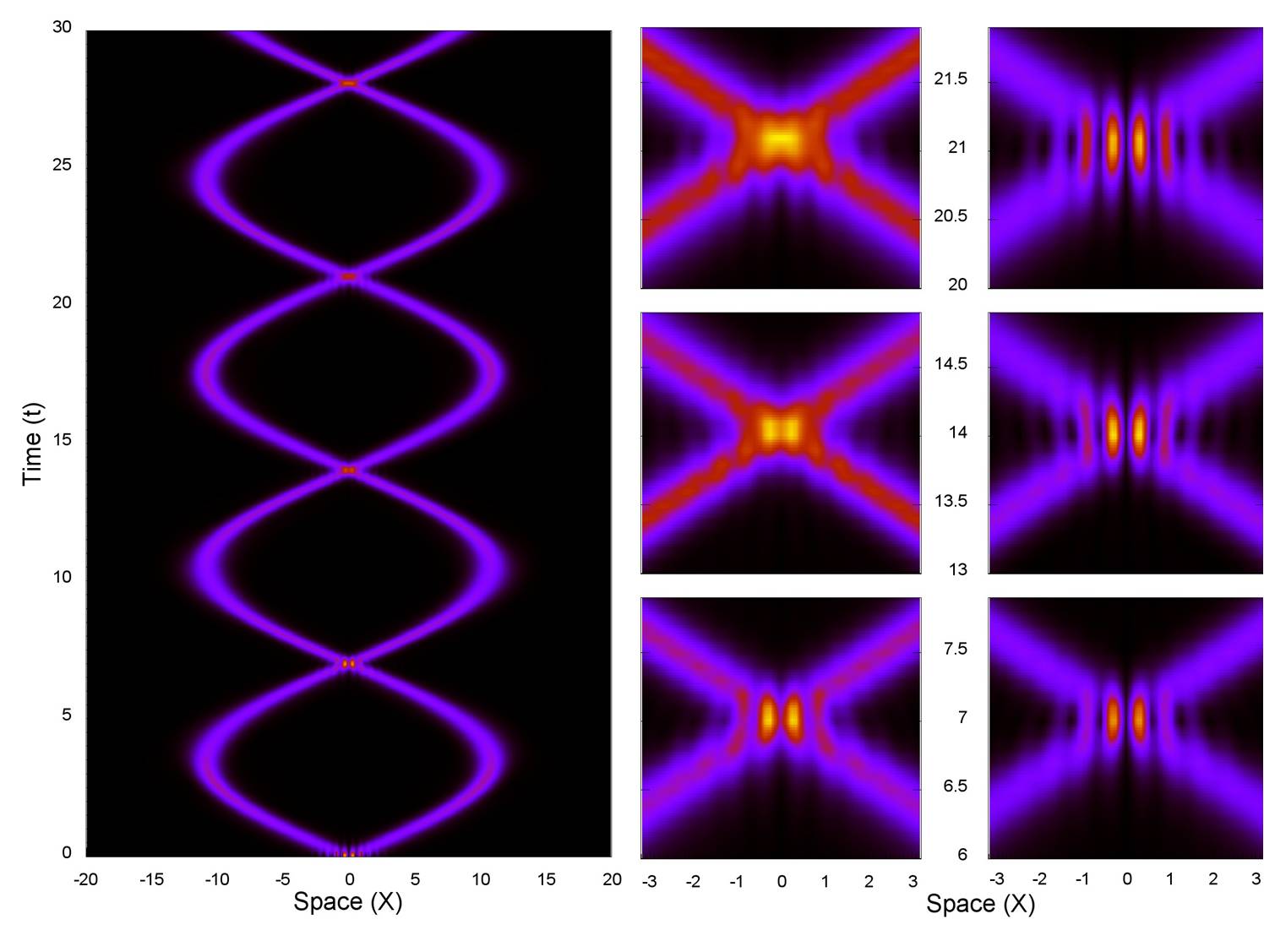}
\caption{(color online). 
The first $\pi$-pulse $e^{i k x}\!+\!e^{-i kx\!-\!i \pi}$ to split the system. 
Evolution of the initially-coherent cloud confined in a weak harmonic trap, $V(x)\!=\!0.1x^2$, 
after a sudden application of the laser-pulse with $k\!=\!5$. 
The initial $sech[x]$-shaped soliton is made of $N\!=\!100$ attractive bosons with $\lambda_0\!=\!-0.04$.
The left panel shows in a space-time representation how the split matter-waves oscillate.
The middle panels show enlarged densities at different re-collision times computed at the many-body level;
for comparison the right panels depict the corresponding Gross-Pitaevskii results.
The many-body computations reveal a degradation of the interference patterns with time
while the mean-field (GP) theory demonstrates ideal $\pi$-phase interferences.
The many-body physics behind is a loss of the coherence due to build-up of the two-fold fragmentation.
All quantities shown are dimensionless.}
    \label{fig01}
\end{figure}

The physics of $N$ low-dimensional bosons trapped in an external potential $V(x)$ 
with a contact inter-boson interaction is governed by the many-body Hamiltonian
$\hat H  \!=\! \sum_{j=1}^N \left[- \frac{1}{2} \nabla^2_{x_j} \!+\! V(x_j,t) \right]
\!+\!\sum_{j < k}^N \lambda_0 \delta(x_j\!-\!x_k) \nonumber$, here written in dimensionless units, $\hbar\!=\!1$, $m\!=\!1$. 
In this work we study attractive inter-particle interactions, i.e., $\lambda_0<0$.
As an initial state for interferometry we use the fully condensed idealized single-hump soliton 
with $sech[\gamma x]$-profile and width defined by $\gamma$  \cite{Gamma_OPT}. 
To contrast the many-body and the mean-field results we use the same initial state.

The first interferometric stage is to split the initially-condensed cloud.
Our main assumption is that the process of splitting by the applied laser pulse
can theoretically be described by a {\it sudden} imprinting of momentum and phase on each and every atom of the cloud,
$\psi(x,t\!=\!0) [e^{ikx}\!+\!e^{-ikx\!-\!i \chi}]$,
where $\psi(x,t\!=\!0)$ is the initial N-boson wave-function, 
$k$ is the imparted momentum, and $\chi$ is the imprinted phase \cite{IMPINT_PSI}.
If the split cloud after the momentum imprinting were to keep the coherence and solitonic properties,
 $\chi$ would define the phase between the split matter-waves.
Namely, $\chi=0$ results in an in-phase (gerade) two-hump soliton and $\chi=\pi$ in an out-phase (ungerade) one.


To determine the region of the imparting momentum leading to a complete split of the solitonic cloud
we propagate the full many-body TDSE for the initial $sech[\gamma x]$-shaped soliton 
multiplied at $t\!=\!0$ by the $e^{i k x}\!+\!e^{\!-i kx\!-\!i\chi_1}$.
The soliton is made of $N\!=\!100$ attractive bosons with $\lambda_0\!=\!-0.04$ and $\gamma\!=\!1$  \cite{Gamma_OPT}.
The density becomes fully split into two sub-clouds propagating in opposite directions
when the applied momenta $k\!>\!4.0$.
For a smaller momentum the split is incomplete because the attractive forces prevent it.
The phase $\chi_1$ does not affect the fast splitting process, but it governs different scenarios of
degradation of the coherence between the split matter-waves, see the supplementary material \cite{SM} for details.
A weak external trap does not impact the fast split process either.

The second interferometric step is to deflect the split parts. 
We use a weak harmonic trap to reflect the propagating sub-clouds back. 
The left panel of Fig.~\ref{fig01} depicts in a space-time representation
the many-body evolution of the initially-coherent solitonic wave-packet with $\gamma=1$ \cite{Gamma_OPT} 
confined in a weak harmonic trap $V(x)\!=\!0.5 \omega^2 x^2\!=\!0.1 x^2$ and
split by the applied pulse with $k\!=\!5$ and $\chi_1\!=\!\pi$.
The many-body simulations confirm the solitonic-like behavior of the studied system -- 
the split sub-clouds continue to oscillate by bouncing from the walls and re-colliding twice during the oscillation period
without significant broadening.
The GP mean-field theory applied to the same initial condition 
reproduces the overall oscillating behavior of the split system, but not the re-collision events.
In the middle panels of Fig.~\ref{fig01} we plot on enlarge scales the densities at the re-collisions computed at the many-body level.
The corresponding GP results are depicted in the right panels for comparison.
The main difference seen is that the many-body results reveal a blurring -- degradation of the fringe visibility,  
indicating on the loss of coherence in the system.

To quantify the coherence we relay on the well-known and  
widely-used definition of condensation given by Penrose and Onsager  \cite{frag0}
in terms of the natural occupation numbers -- the eigenvalues $n_i$ of the reduced one-particle density matrix \cite{DNS}.
When only a single eigenvalue $n_1$ has macroscopic occupation the system is condensed, when several -- it is fragmented. 
These definitions of condensation and fragmentation are valid at zero temperature,
for finite-N systems with general inter-particle interactions in any dimension.
They are also applicable for time-evolving systems, e.g., for the non-stationary processes studied here.

The initial state used in our simulations is fully condensed with all 100\% bosons residing in a single ``solitonic" natural orbital.
The GP mean-field theory is incapable to describe the loss of coherence
and keeps $n_1=100\%$ condensation during all the propagation times, irrespective to the value of the initial phase $\chi_1$.
The many-body treatment significantly modifies this idealized mean-field picture --
after the applied pulse, an initially unoccupied second natural orbital becomes macroscopically populated with time, see \cite{SM} for details.
Indeed, at the moments when the split matter-waves depicted in the middle panels of Fig.~\ref{fig01} re-collide for the 
first, second, and third times the occupation of the second natural orbital (fragment)  
becomes $n_2=26.4\%$,43.7\%, and 45.7\%, respectively.
Comparing  these occupations and the corresponding interference patterns depicted 
in the middle panels of Fig.~\ref{fig01}, one can conclude that the closer $n_2$ is to 50\%, the worse is the visibility of the fringes.
From now on we use $n_2$ to quantify the loss of coherence via the fragmentation.
It is worth mentioning that for the first applied pulse with $\pi$-phase, 
the most occupied (first) natural orbital after the split has a profile 
with two out-of-phase (ungerade) humps. The second natural orbital, in order to keep the mutual orthogonality with respect to the first one, 
has to have gerade structure, i.e.,  0-phase between the constituting humps.

\begin{figure}[t]
\vspace*{0cm}
\includegraphics[width=8cm ,angle=0]{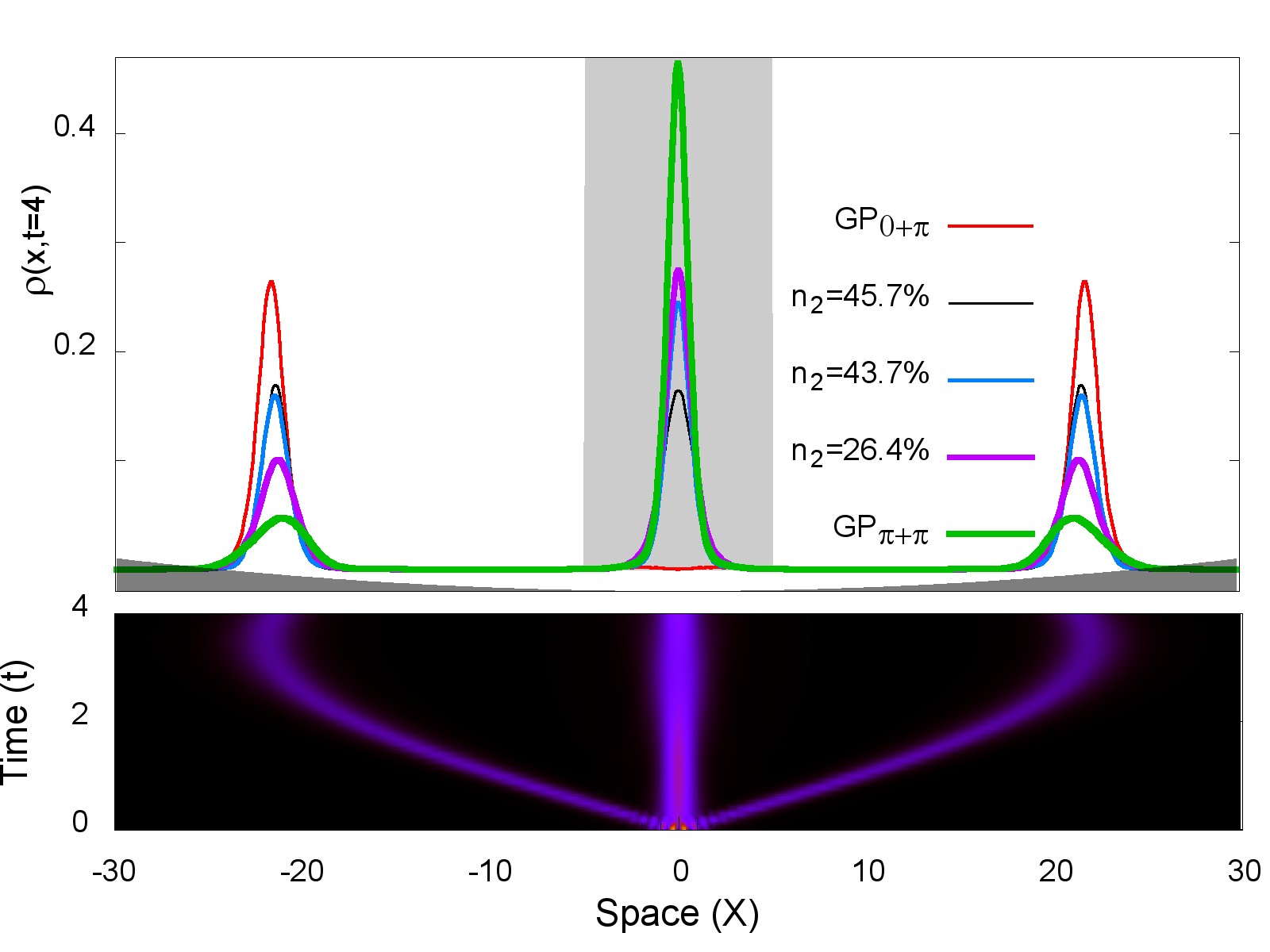}
\caption{(color online).
The second $\pi$-pulse $e^{i k x}\!+\!e^{-i kx\!-\!i \pi}$ to measure fragmentation.
By applying a second laser-pulse with $k\!=\!5$ to the system from Fig.~\ref{fig01},
at the moment when the two previously split sub-clouds re-collide, results 
in three matter-waves propagating with different momenta $-2k,0k,+2k$ as depicted in the lower panel.
The relative populations of these sub-clouds ($k-$channels) depend on the fragmentation of the state at the re-collision moment.
The occupation of the second fragment (natural orbital) $n_2$
is proportional to the number of $0k-$ atoms in the central (gray) area, also called visibility $\nu$, see Eq.~\ref{Nu_n2}.
So, one can directly access the non-condensed, i.e., fragmented fraction $n_2$ in the system 
-- this is the main result of the present study.
All quantities shown are dimensionless.}
    \label{fig02}
\end{figure}

Direct registration of the visibility of interference fringes in the density of the re-colliding attractive system as, e.g., shown in Fig.~\ref{fig01}
would unambiguously prove whether the multi-hump structures seen in the majority of experiments with attractive BECs
are coherent as believed for a long time or rather fragmented as our many-body simulations indicate, see also Refs. ~\cite{fragmenton, swift}.
However, due to small sizes of the attractive (sub)clouds it is technically very difficult to detect the interferences directly.
Fortunately, the phase between the split sub-clouds, as well as the contrast of the interference pattern,
can be determined accurately within the third interferometric step -- by applying 
a recombining (second) laser-pulse exactly at the re-collision moment, and monitoring the populations of the momentum channels.

Let us continue the interferometric protocol with the system from Fig.~\ref{fig01}, first at the mean-field level.
To simulate the recombining interferometric step we apply a second laser-pulse $e^{i k x}\!+\!e^{-i kx\!-\!i\chi_2}$ 
at the moments when the two previously-split sub-clouds re-collide.
At the idealized mean-field (GP) level and if we neglect the interaction during the splitting and re-collision(s),
these time-moments in a harmonic trap with confining frequency $\omega$ are exactly known -- 
$t^{rc}_{j}\!=j \pi/\omega$ with $j=1,2,3$ corresponding to the first, second, and third re-collision events.
Moreover, in this approximation the wave-functions at the re-collisions are the same as the original wave-packet at $t\!=\!0$.
So, the result of imprinting of a second laser-pulse at the re-collision event is equivalent to a multiplication of the  
two terms: $\psi(x,t\!=\!0) [e^{i k x}\!+\!e^{-i kx\!-\!i \chi_1}] [e^{i k x}\!+\!e^{-i kx\!-\!i \chi_2}]$.
Simple algebra shows that application of two identical pulses $\chi_1\!=\!\chi_2\!=\!\pi$
would result in splitting of the original cloud into three parts:
$\psi(x,t\!=\!0) [e^{i k x}\!-\!e^{-i kx}] [e^{i k x}\!-\!e^{-i kx}]\equiv \psi(x,t\!=\!0)[e^{i 2k x}\!-\!2\!+\!e^{-i 2kx}]$.
The left and right matter-waves are propagating with the doubled momenta in opposite directions,
while the middle one stays at rest and has a zero momentum. 
The corresponding numerical GP simulation depicted in the lower panel of Fig.~\ref{fig02} confirms it.
Here we take the GP wave-packet from the previous interferometric run (see the right panels of Fig.~\ref{fig01} at $t^{rc}_1\!\approx\!7.0$)
and apply to it the second pulse identical to the first one, i.e,  with $\chi_2\equiv\chi_1=\pi$ and $k=5$. 
The cut of the space-time density at $t=4.0$ where all three matter-waves 
are separated in space is plotted in the upper panel of Fig.~\ref{fig02} and marked as $GP_{\pi+\pi}$. 
Similarly, when two applied pulses have maximally different phases, say $\chi_1\!=\!0$, $\chi_2\!=\!\pi$
-- one would observe only two sub-clouds traveling with $\pm2k$ momenta in opposite directions:
$\psi(x,t\!=\!0)[e^{i k x}\!+\!e^{-i kx}][e^{i k x}\!-\!e^{-i kx}] \equiv \psi(x,t\!=\!0)[e^{i 2k x}\!-\!e^{-i 2kx}]$.
We have performed the GP simulation for this scenario as well and plot the computed density 
at $t=4.0$ marked as $GP_{0+\pi}$ in the upper panel of Fig.~\ref{fig02}.
The density at the trap's origin is hardly present  confirming thereby that in this case 
the first applied pulse creates in-phase (gerade) two-hump soliton, while the second pulse speeds both contra-propagating parts up.

In experiments, after the second applied pulse, when all three matter-waves are separated in space 
one can directly measure the amount of atoms in the sub-clouds, 
i.e., the populations of $-2k,\!0k,\!+2k$-channels ($N^{\!-2k}$:$N^{\!0k}$:$N^{\!+2k}$).
To quantify the visibility of the interference fringes one can use \cite{Interferometry_Soliton_exp1}
the relative ratio $\nu\!=\!{N^{0k}}\!/\!N$, where $N\!=\!N^{-2k}\!+\!N^{0k}\!+\!N^{+2k}$.
For the above-discussed mean-filed results the maximal visibility  $\nu^{max}\!=\!2/3$ is observed when
both applied pulses have the same $\pi$-phase. This corresponds to (1/6:4/6:1/6) populations of the k-channels, see $GP_{\pi+\pi}$ density in
the Fig.~\ref{fig02}.
The minimal visibility, $\nu^{min}\!=\!0$, corresponds to (1/2:0:1/2) population of the k-channels and is
obtained when the applied pulses have different phases, see $GP_{0+\pi}$ density in Fig.~\ref{fig02}.

Now let us consider what one would expect from the last interferometric pulse
when the re-colliding matter-waves are fragmented, i.e., uncorrelated and incoherent.
An ideal, fully two-fold $n_1=n_2=N/2$  fragmented object  -- the fragmenton  ~\cite{fragmenton} --  can be constructed 
from the coherent symmetric (gerade) and anti-symmetric (ungerade) two-hump solitons, with well-separated humps, 
combined into the  $|N/2,N/2\rangle$ Fock state.
In Ref.~\cite{fragmenton,swift} it was shown that such a combination of solitons lowers the total energy of the combined object
and governs the fast development of fragmentation in initially-coherent multi-hump systems.
If one would enforce a re-collision of the humps of this fragmented Fock state in a harmonic trap 
and apply the recombining laser-pulse of an appropriate momentum at the re-collision event, 
one would observe the (1/3:1/3:1/3) populations of the k-channels. 
This conclusion comes directly from the above-discussed analysis of the laser-pulses applied to the fully-coherent two-hump solitons.
Indeed, when the recombining $\pi$-phase pulse is applied to the gerade two-hump solitonic part 
constituting the $|N/2,N/2\rangle$ state, two sub-clouds traveling with $\pm2k$ momenta are formed.
The contribution from the complimentary orthogonal part  which is the ungerade soliton provides us with three matter-waves
propagating with $0k$- and $\pm2k$ momenta. Consequently, the combined relative occupations of the k-channels
for the $|N/2,N/2\rangle$ state become equal.
The mutual orthogonality of the constituting gerade and ungerade parts allows us 
to apply the same logic to a general Fock state $|n_1,n_2\rangle$, $N=n_1+n_2$.
In such a two-fold fragmented system the occupation of the second natural orbital $n_2$ and 
the population of the $0k$-channel (the visibility $\nu$) are linearly connected:
\beq
\label{Nu_n2}
n^{\mathit{intf}}_2/N=1-\frac{3}{2}\nu.
\eeq
Here the subscript $\mathit{intf}$ is used to distinguish the interferometric measurement of the fragmentation from the exact value 
available after the diagonalization of the one-body density \cite{DNS}.

The above relation, Eq.~(\ref{Nu_n2}), has been obtained for the idealized single-configurational $|n_1,n_2\rangle$ Fock state
built from the symmetric and anti-symmetric orbitals.
So, it is also valid in a many-body realm where the propagating system 
is described by a linear combination of these Fock states $\sum_jC_j|N-j,j\rangle$ with different weights $C_j$.
To demonstrate it we study the last interferometric step within the split scenario from Fig.~\ref{fig01}, now at the many-body level.
The recombining laser-pulse with $\pi$-phase and $k=5$ applied to the many-body wave-function 
taken from the middle panel of Fig.~\ref{fig01}, at the first re-collision event $t^{rc}_1\!\approx\!7.0$, splits it into three sub-clouds.
At this moment the system is momentarily fragmented with $n_2\approx 26.4\%$ \cite{DNS}.
After the applied pulse we propagate the system further till $t=4.0$ where all three matter-waves are separated in space,
the computed density marked as $n_2=26.4\%$ is depicted in the upper panel of Fig.~\ref{fig02}.
By integrating it in the gray area, corresponding to the $0k$-channel, one gets $\nu=0.4889$.
From Eq.~(\ref{Nu_n2}) we obtain the interferometric fragmentation $n^{\mathit{intf}}_2\approx26.7\%$.
We apply the same protocol to the many-body states taken from Fig.~\ref{fig01} at the second $t^{rc}_2\!\approx\!14.05$
and third $t^{rc}_3\!\approx\!21.08$ re-collisions. 
The computed densities after the separation are marked in the upper panel of Fig.~\ref{fig02} as  $n_2=43.7\%$ and $n_2=45.7\%$
according to the actual occupation of the second natural orbital at the respective re-collisions.
The integrations of the $0k$-areas gives the interferometric fragmentation $n^{\mathit{intf}}_2\approx42.8\%$ ($\nu=0.3817$)
for the second re-collision and  $n^{\mathit{intf}}_2\approx45.0\%$ ($\nu=0.3668$) for the third one.
In all these examples the $\nu$-based interferometric fragmentation $n^{\mathit{intf}}_2$ 
is very close to the corresponding natural occupation number obtained by the diagonalization \cite{DNS}.
Small deviations are mainly due to the mismatch in the determination of the re-collision times, see \cite{SM} for details.
Concluding, the above-discussed interferometric protocol and Eq.~(\ref{Nu_n2}) allows one for 
the direct measurements of the fragmentation.

In this letter we have investigated interferometry in low-dimensional ultra-cold systems with attractive inter-particle interaction 
at the many-body level. Two matter-waves obtained from the initially-coherent cloud by an applied laser-pulse
inevitably loose the inter-hump coherence and become fragmented due to inter-particle attractions.
The degree of fragmentation can be directly measured by re-colliding the constituting sub-clouds 
with a simultaneous imparting of a proper momentum at the re-collision time.
Within this protocol fragmentation becomes directly proportional to the occupation of the zero-momentum channel.
Furthermore and without loss of generality, the proposed protocol to discriminate coherent and fragmented many-body systems with two-hump densities 
can be expanded to measure fragmentation in generic many-body systems with attractive, repulsive short- and long-range inter-particle interactions.

We hope this work will stimulate the direct experimental measurements of fragmentation in attractive BECs with 
two-hump densities and validate the predicted interaction-induced loss of inter-hump coherence. This knowledge can help
to improve matter-wave interferometry further.

\subsection*{Acknowledgments} 
Computation time on the HLRS, bwGRiD, and HybriLIT clusters as well as financial support by the DFG are greatly acknowledged.

\newpage

\renewcommand{\figurename}{Figure S\hglue -0.12 truecm}

\onecolumngrid

\section*{\large Supplemental Material\\ $ \ $ \\Interferometry with correlated matter-waves}


In this supplementary material we show: 
(i) how to apply the procedure of sudden imprinting of momentum by a laser-pulse at the many-body level 
within the multiconfigurational time-dependent Hartree for bosons (MCTDHB) method;
(ii) how the phase of the split-pulse applied to the initially-coherent soliton governs the loss of the inter-hump coherence;
(iii) how the fringe visibilities depend on the phase difference between the split and recombining pulses.

\section*{How to imprint momentum in the MCTDHB method}

The Multi-Configurational Time-Dependent Hartree method for Bosons (MCTDHB)
is used to solve the time-dependent Schr\"odinger equation:
$$
 \hat H \Psi = i \frac{\partial \Psi}{\partial t}.
$$
Here $\hbar=1$ and $\hat H$ is the generic many-body Hamiltonian defined in the main text.
The MCTDHB {\it ansatz} for the many-body wave-function $\Psi(t)$ is taken as a linear combination of time-dependent permanents:
$$
\left|\Psi(t)\right> =
\sum_{\vec{n}}C_{\vec{n}}(t)\left|\vec{n};t\right> \, ,  
$$
where the summation runs over all possible configurations whose
occupations $\vec{n}= (n_1,n_2,n_3,...,n_M)$ preserve the total number of bosons $N$.
So, the Fock state $|n_1,n_2\rangle$ studied in the main text is one of the terms in this sum.
The expansion coefficients $\{C_{\vec{n}}(t)\}$ and shapes of the orbitals $\{\phi_j(x,t)\}^M_{j=1}$ 
are variational time-dependent parameters of the MCTDHB method,
determined by the time-dependent variational principle.

To solve a time-dependent problem means to specify an initial many-body state and to find how the many-body wave-function evolves.
The initial MCTDHB many-body state is given by initial expansion coefficients $\{C_{\vec{n}}(t=0)\}$ and by initial shapes of the orbitals $\{\phi_j(x,t=0)\}^M_{j=1}$.
The famous Gross-Pitaevskii (GP) mean-field is a limiting case of the MCTDHB ansatz, obtained when only a single orbital, i.e., $M=1$,
enters the expansion. In this case only one configuration $|N\rangle$ with the $\phi_{GP}(x,t=0)$ orbital is needed, $C_{|N\rangle}\equiv1$.

Here it  is worthwhile to recall that the many-body MCTDHB(M) wave-function has the following general property (freedom):
any unitary operation applied to the orbitals $\{\phi_j(x,t)\}^M_{j=1}$, e.g., 
a linear combination representing a rotation between symmetric and localized orbitals, 
can be compensated by a "contra-rotation" of the expansion coefficients $\{C_{\vec{n}}(t)\}$, 
keeping all the observables and, thereby, the physics  unchanged.

At the MCTDHB$(1)\equiv$GP level the procedure of imprinting of the momentum $k$ in one-dimensional setups
is well-defined -- one multiplies the GP orbital $\{\phi_{GP}(x,t=0)\}$ by the respective function: 
$$
\tilde{\phi}_{GP}(x,t=0)=\phi_{GP}(x,t=0) exp(-i k x).
$$
The application to a more general case is straightforward, e.g., $\phi_{GP}(x,t=0)( e^{i k x}+2+e^{-i k x})$.
The resulting orbital has to be properly re-normalized. 

In the general MCTDHB(M) case the situation is more delicate.
In the one dimensional case, to imprint a momentum $k$ to each and every atom of the system at $t=0$ 
one can use the same receipt as in the GP case --
the initial shapes of the orthonormal orbitals $\phi_j(x,t=0)$ are multiplied by the $exp(-i k x)$ prefactor (function):
$$
\tilde{\phi_j}(x,t=0)=\phi_j(x,t=0) exp(-i k x), \,\, j=1,...,M.
$$
This transformation keeps the resulting orbitals orthogonal for any set of the initial orthonormal orbitals.

For more sophisticated pulses a similar approach is applicable, but
one should take care about the orthogonality and normalization of the transformed orbitals.
For example, the $exp(i k x)+exp(-i k x)$ pulse can be applied to the MCTDHB(2) wave-function 
only when the respective orbitals have symmetric (gerade) and anti-symmetric (ungerade) structure.
For a localized (rotated) set of the orbitals such a transformation is inapplicable, 
because the transformed orbitals become non-orthogonal:
$$
\tilde{\phi_j}(x,t=0)=\phi_j(x,t=0)( e^{i k x}+e^{-i k x}) ,\, j=1,2,
$$
$$
\langle \phi_1(x,t=0)( e^{-i k x}+e^{i k x}) |  \phi_2(x,t=0)( e^{i k x}+e^{-i k x}) \rangle= 
$$
$$
\langle \phi_1(x,t=0)|\phi_2(x,t=0)( 2+2 (e^{i 2 k x}+e^{-i 2 k x}) \rangle=\langle \phi_1(x,t=0)|\phi_2(x,t=0) 2 cos( 2 k x)) \rangle.
$$
The last integral can be zero if the orbitals have different symmetries or a zero overlap.
Of course, the orthogonality of the orbitals after the applied pulse 
can be restored manually by applying the Gram-Schmidt renormalization technique.
An alternative and more consistent way is to apply a unitary operation rotating the initial orbitals and respective expansion coefficients
before the applied pulse such that the symmetry of the orbitals is restored. 
For both $\pi$-pulses investigated in the main text the symmetry of the orbitals is naturally present and preserved.
For the general case, discussed below, where the second applied pulse has a different phase,
we have used the above discussed Gram-Schmidt re-normalization.

\section*{Degradation of the coherence and development of fragmentation}

\begin{figure}[t]
\vspace*{0cm}
\includegraphics[width=11cm ,angle=0]{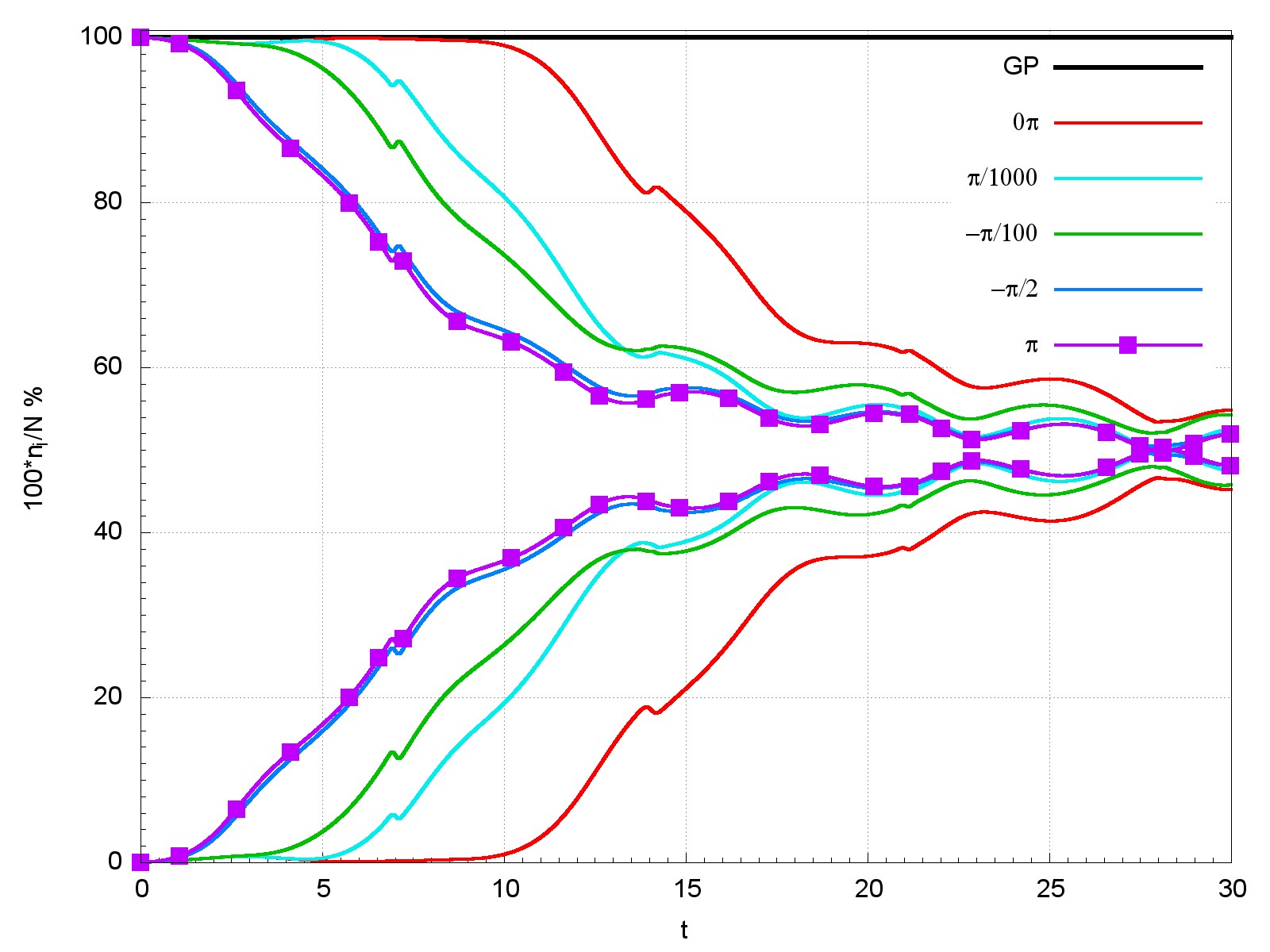}
\caption{(color online).
Dynamics of the loss of coherence for several different initial phases $\chi_1$ 
of the split-pulse $e^{i k x}\!+\!e^{-i kx\!-i\!\chi_1}$ applied to the fully condensed (coherent) idealized $sech[x]$-shaped soliton. 
Evolutions of the first  $n_1$ and  second $n_2$ natural occupation numbers 
obtained as eigenvalues of the reduced one-particle density matrix at each propagation time are plotted.
After the applied split-pulse the second natural orbital quickly becomes populated.
Comparing the interference patterns plotted in middle panels of Fig.~1 
and the values of $n_2$ depicted by line-point-style curve for the same system, one concludes that
the development of the two-fold fragmentation is accompanying by degradation of the fringe visibility.
In contrast, the Gross-Pitaevskii theory predicts 
complete coherence of the system for all the propagation times and for all the applied phases.
All quantities shown are dimensionless.}
    \label{fig01sup}
\end{figure}

We have applied the $exp(i k x)+exp(-i k x - i\chi_1)$ pulse with $k=5$ to the same initially-coherent idealized 
$sech[x]$-shaped soliton $|N,0\rangle$ as in the main text. 
For this initial many-body state it is legitimate (technically) to apply pulses of any shape,
because the second involved orbital has zero occupation.
After the applied pulse irrespective to its phase $\chi_1$, the initially unoccupied second natural orbital' occupation $n_2$ 
becomes macroscopically populated with time, indicating on the fragmentation in the system, see Fig.~S\ref{fig01sup}.
The results depicted in this figure demonstrate that the coherence between 
the split sub-clouds is lost via the development of macroscopic two-fold fragmentation.
Interestingly, the way how the coherence is lost does depend on the imprinted initial phase $\chi_1$.
For the phase $\chi_1\!=\!0$ the coherence is ``preserved'' during the longest time, while already a
very small $\pi/1000$ phase results in a fast degradation of the coherence.
From a practical point of view it means that an attempt of experimental stabilization 
of the coherence by adjusting the phase of the split laser pulse would not lead to significant improvements.
Here it is worthwhile to mention that similar simulations done with a more ``realistic'' initial state obtained 
as a many-body ground state of the same system with attractive interaction in an harmonic $V(x)=0.1x^2$ trap, 
gives very similar results, implying that a small (less a percent) 
depletion of the initial cloud does not change the physics -- after the spit  the coherence is lost.

\section*{The general case, when the split and recombining pulses have different phases}

\begin{figure}[t]
\vspace*{0cm}
\includegraphics[width=11cm ,angle=0]{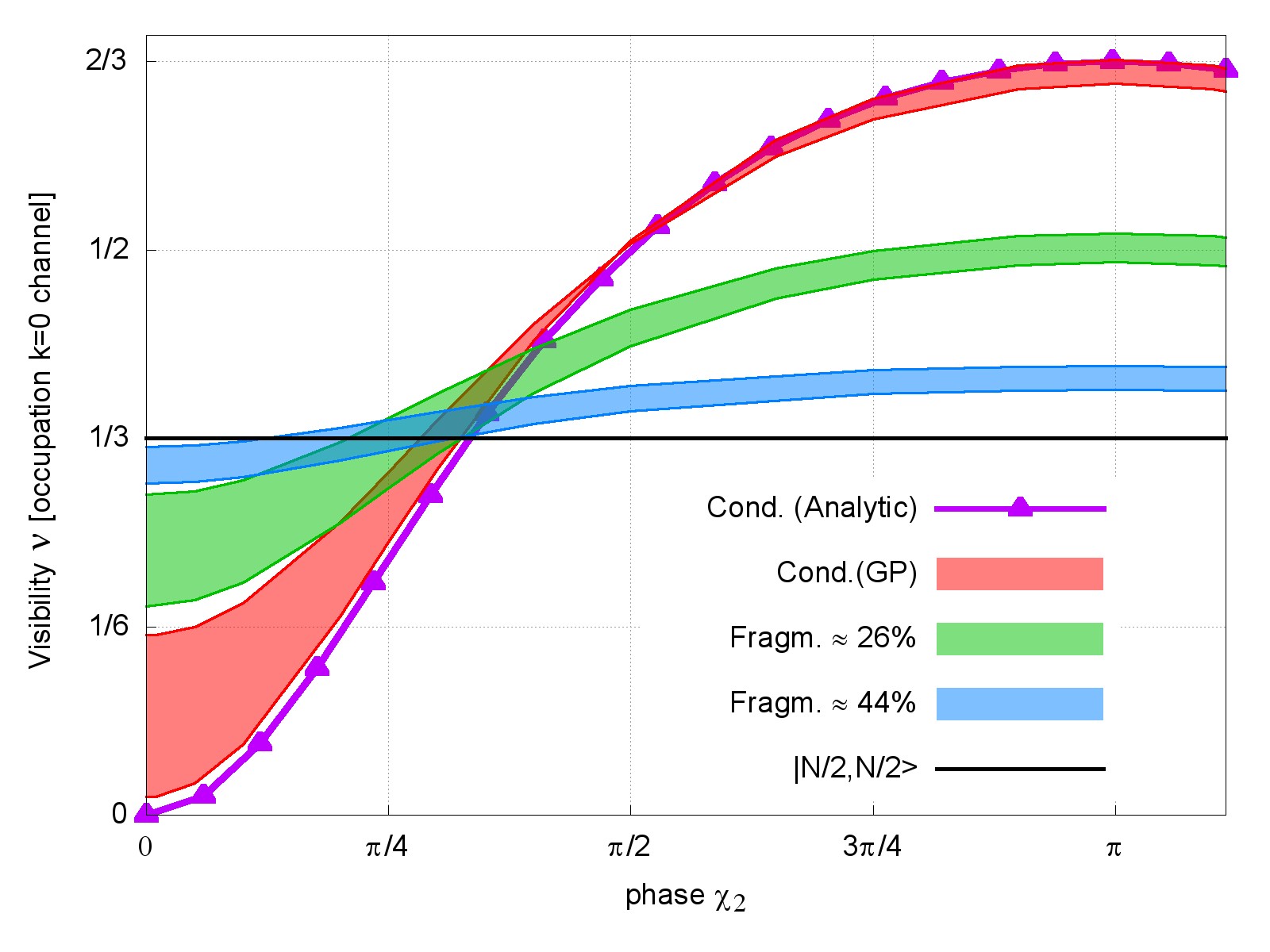}
\caption{(color online). Measurement of fragmentation by monitoring the visibility $\nu$: 
Impact of the relative phase on the population of $0k$-channel.
The application of the second laser-pulse $e^{i k x}\!+\!e^{-i kx\!-\!i \chi_2}$ to the previously $\pi$-split system (see Fig.~1 of the main text)
re-colliding in a weak harmonic trap $V(x)\!=\!0.1x^2$  leads to 
formation of three matter-waves propagating with different momenta $-2k,0k,+2k$. 
The relative populations of these sub-clouds ($k-$channels) depend on the fragmentation of the state at the re-collision moment.
For an ideal condensed mean-field state, previously split by the $\pi$-pulse and remaining fully condensed afterwards, 
the visibility as a function of $\chi_2$ oscillates  as $\nu\!=\![1-cos(\chi_2)]/[2-cos(\chi_2)]$, see curve with triangles.
The numerical GP simulations done for this scenario at slightly mismatched re-collision times $t^{rc}_1\!\in\![7.0:7.1]$ 
are depicted by the red filled area.
An ideal fully-fragmented state $|N/2,N/2\rangle$ has a phase-independent visibility $\nu\!=\!1/3$.
The many-body systems with finite fragmentation $n_2\!\approx\! 26.4\%$ and $n_2\!\approx \!43.7\%$ (see Fig.~S\ref{fig01sup})
have visibilities oscillating around this limiting value, the corresponding results are depicted by the green and blue areas.
The maximal population of the $0k-$channel, i.e., $\nu^{max}$ here reached for $\chi_1=\chi_2=\pi$ 
is proportional to the occupation $n_2$ of  the second natural orbital:
$n^{\mathit{intf}}_2/N\!=1-\!\frac{3}{2}\nu^{max}$. By computing this maximal visibility 
one can measure the fragmentation in the system -- this is the main result of the present study.
All quantities shown are dimensionless.}
 \label{fig03}
\end{figure}

Here we simulate the full interferometric sequence within the two-pulses protocol used in the main text
for the case when the applied pulses have different phases.
Specifically, we have applied it for the same system under investigation
and used the following protocol (with the above caution on the re-enforcement of the  ortho-normalization of the involved orbitals):
(i) the ideal soliton is split by the $\pi$-phase pulse $e^{i k x}\!+\!e^{-i kx\!-\!i \pi}$ 
and propagated in the trap (see Fig.~1) within the MCTDHB equations
till the first re-collision event at $t^{rc}_1\!\approx\!7.0$  where the system becomes $n_2\approx 26.4\%$ fragmented;
(ii) at the re-collision time a second recombining laser-pulse of the same momentum $k=5$ is applied;
(iii) we propagate the MCTDHB equations next $\!4$ units of time till the full separation of the sub-clouds, and measure 
the visibility $\nu$ which equals to the relative occupation of the 0k-channel. 
Next, we repeat steps (ii) and (iii) for different phases $\chi_2$ of the second applied pulse. 
The computed visibility as a function of the phase of the second pulse applied  to the systems at the first re-collision event
is depicted in Fig.~S\ref{fig03} by the green area.
We have also applied  the same protocol and the steps (ii) and (iii) to the $n_2\approx 43.7\%$ fragmented system 
taken from Fig.~1 at the second re-collision event $t^{rc}_2\!\approx\!14.05$.
The computed visibility is depicted by the blue area.
These areas correspond to possible mismatch (uncertainty) in determination of the moments of the first $t^{rc}_1\in [7.0:7.1]$ 
and second $t^{rc}_2\in[14.0:14.1]$ re-collision events.
From Fig.~S\ref{fig03} it is clearly seen that the closer the system to an ideal 50\%-50\% 
fragmented state is, the smaller is the amplitude of the oscillation of the visibility. 

The same two-pulse protocol simulated within the GP theory
provides us with a perfect oscillation of the visibility $\nu\!=\![1\!-\!cos(\chi_2)]/[2\!-\!cos(\chi_2)]$
as a function of the phase $\chi_2$ of the second applied pulse.
This analytical formula correctly reproduces both limiting cases discussed in the main text.
The minimal  visibility $\nu^{min}\!=\!0$ is observed when the pulses have maximally different phases $\chi_1=\pi$ and $\chi_2=0$, resulting in (1/2:0:1/2) occupations of the k channels. The maximal visibility  $\nu^{max}\!=\!2/3$ is observed 
when the both pulses have the same ($\chi_1=\chi_2=\pi$) phases, resulting in (1/6:4/6:1/6) populations.

\begin{thebibliography}{64}
\bibitem{Gamma_OPT} 
Within the Gross-Pitaevskii theory the ground-state solution of the studied Hamiltonian with $V(x)\!=\!0$
has the optimal width $\gamma=|\lambda_0|(N-1)/2$.
The optimal width of the two-hump soliton with well-separated humps is $\gamma=|\lambda_0|(N-1)/4$.
In this work to study coherence in two-hump objects made of $N=100$ bosons and $\lambda_0=-0.04$ 
we use $\gamma=1.0$.
\bibitem{IMPINT_PSI} We use the following shorthand notation:
$\psi(x,t\!=\!0) [e^{ikx}\!+\!e^{-ikx\!-\!i \chi}]\equiv \Psi(x_1,x_2,...,x_N,t\!=\!0) \prod^N_{j=1} [e^{ikx_j}\!+\!e^{-ikx_j\!-\!i \chi}]$.
\bibitem{DNS} For a given many-body wave function $\Psi(x_1,\ldots,x_N;t)$ we construct 
density:
$\rho(x_1,x'_1;t) =  N\int \Psi^\ast(x'_1,x_2,\ldots,x_N;t) \Psi(x_1,x_2,\ldots,x_N;t) dx_2\cdots dx_N$
and find its eigenvalues $n_k$ and eigenvectors 
$\rho(x,x';t)=\sum^M_{k=1} n_k(t) \phi_k^{\ast NO}(x',t) \phi_k^{NO}(x,t)$.
\end{thebibliography}
\end{document}